\documentclass{IOS-Book-Article}

\usepackage{mathptmx}
\usepackage{mathtools}

\usepackage{amssymb}
\usepackage{amsfonts}
\usepackage{amsmath}
\usepackage{graphicx}
\usepackage{epsfig}
\usepackage{color}
\usepackage{subfig}
\usepackage{epstopdf}
\usepackage{algorithm}
\usepackage{algpseudocode} 
\makeatletter
\def\BState{\State\hskip-\ALG@thistlm}
\makeatother

\newtheorem{lem}{Lemma}[section]

\newtheorem{cor}{Corollary}[section]

\def\hb{\hbox to 10.7 cm{}}

\begin{document}

\pagestyle{headings}
\def\thepage{}

\begin{frontmatter}              

\title{{\LARGE \textbf{Toward Intelligent Traffic Light Control with Quality-of-Service Provisioning}}}


\author[A]{\fnms{Lei Miao}%
\thanks{Corresponding Author: Box 19, Middle Tennessee State University, 1301 E Main St, Murfreesboro, TN 37132,  USA, E-mail:
lei.miao@mtsu.edu .}},
\author[B]{\fnms{Lijian Xu}}

\address[A]{Mechatronics Engineering, Middle Tennessee State University}
\address[B]{Dept. of Electrical and Computer Engineering Technology, Farmingdale State College}

\begin{abstract}
Today’s fixed-cycle traffic signaling is highly suboptimal and aggravates traffic congestion and waste of energy in urban areas. In addition, it offers no quality-of-service guarantee and makes travel time prediction extremely hard. While existing traffic light control research primarily focuses on improving the average wait time of cars, we study in this paper how traffic light scheduling affects the worst-case wait time. In particular, we derive the time a car spends at an intersection in the best-case and the worst-case, respectively. Using the theoretical results, we propose a simple but effective controller and run simulation to verify its performance. The result shows that it works much better than fixed-cycle controllers in both light and heavy traffic scenarios. 
\end{abstract}

\end{frontmatter}

\section{Introduction}

Traffic congestion is a serious problem in many urban areas all over the world and has become a severe challenge to the sustainability of economic grow and urbanization. A report from the Texas Transportation Institute says that in 2011, traffic congestion costs drivers \$121 billion in the U.S. Americans also spent 5.5 billion additional hours sitting in traffic in the same year. That comes down to \$818 per commuter in lost time and wasted fuel. Traffic congestion, to a large extent, is caused by the inefficiency of today’s traffic signal systems, which primarily use fixed-cycle scheduling algorithms. Fixed-cycle approaches are derived off-line based on historical traffic data. The major drawback of fixed-cycle approaches is that they cannot adapt to traffic fluctuation in real-time and may aggravate the congestion when the traffic pattern changes. Therefore, fixed-cycle traffic signaling is highly suboptimal and needs to be replaced with a better approach.

Another problem of today's traffic control system is that it does not provide any quality-of-service (QoS) guarantee. For example, it cannot ensure that each car only spends up to certain amount of time at an intersection. As a result, it is hard to estimate how long it would take a car to travel from point A to point B; this is especially true in case of traffic jam. 

Transportation systems are analogous to communication networks in many aspects: cars travel from one place to another whereas packets are sent from one device to another; roads are like communication links where packets are transmitted in a first-come-first-served fashion; and intersections are very much similar to routers and switches. A question then naturally arises: \textit{why is QoS readily available in communication networks but very lacking in transportation systems?} The answers to it are multi-fold, but the major reason is that historically, transportation systems simply do not have the necessary information of the cars, such as where and how fast they are and where they are headed.

Fortunately, things have changed dramatically in the past ten years or so: most drivers now have smart phones which are capable of reporting the GPS locations and velocities of the vehicles; inexpensive and embedded electronics with similar functionalities are also being built into automobiles; and cars are becoming autonomous and have started to have wireless communication capabilities. All these evidences indicate that transportation systems are ready to be smarter and provide QoS provisioning to travelers. Yet, another complication is that traffic systems contain both time-driven and event-driven dynamics: cars continuously accelerate and decelerate and have to obey Newton's law; and traffic lights switch between green and red at discrete time instances. The hybrid nature makes it hard to model and control in applications related to intelligent transportation systems.

In this paper, we present some preliminary work regarding how a traffic intersection can provide QoS to the cars using it. The contributions of the paper include: \textit{(i)} Different from existing work in the literature, we incorporate car dynamics into our analysis and derive the best-case and worst-case deadline information for a simple two-way intersection; \textit{(ii)} We propose a simple but effective open-loop controller that utilizes the vehicle information to perform control; and\textit{ (iii)} we run simulation to verify the effectiveness and improvement of our approach over the existing fixed-cycle traffic signaling.

The organization of the paper is as follows: in Section II, we summarize
related work; in Section III, we introduce our system model; in Section IV, we present the main results; in Section V, we show simulation results; finally, we conclude and discuss future work in Section VI.

\section{Related Work \label{related_work}}
Perturbation Analysis (PA) and stochastic fluid models \cite{panayiotou2005online} \cite{geng2012traffic} \cite{geng2013quasi} \cite{geng2013multi} are used to find the performance sensitivity with respect to certain parameters of traffic light scheduling. Combined with gradient-based algorithms, such approaches are able to iteratively converge to the optimal solution in an on-line setting. Its powerfulness lies in the fact that it does not rely on the stochastic assumptions of the vehicles' arrival and departure process; in addition, the PA estimator requires very light computation such as counting the number of traffic light switching between certain events.
Fuzzy logic controllers are proposed in \cite{pappis1977fuzzy} \cite{murat2005fuzzy} \cite{collotta2015novel}, in which the theory of fuzzy sets and linguistic control are used to model traffic intersections where analytical methods are in general lacking. Li et al. formulate a traffic light control problem that reduces the stop-and-go time and $co_{2}$ emissions \cite{li2012open}. They achieve this goal by using a three-tier structure as well as a branch-and-bound based real-time algorithm. Ahmad et al. \cite{ahmad2014earliest} considered a scenario that each vehicle has its specific deadline that needs to be satisfied. The performance of two algorithms: Earliest Deadline First (EDF) and Fixed Priority (FP), in terms of the average delay, is compared with that of fixed-cycle scheduling. Reinforcement learning are used in \cite{prashanth2011reinforcement} and \cite{el2013multiagent} to adaptively reduce the average cost. Mixed integer linear programming is used in  \cite{dujardin2011multiobjective} to perform adaptive control on an isolated intersection. In \cite{khamis2012adaptive}, Khamis et al. use Bayesian probability interpretation and the Intelligent Driver Model (IDM) to reduce the average trip waiting time. In \cite{prashanth2012threshold}, the authors use thresholds of the vehicle waiting queues to adjust the traffic light scheduling; specifically, they use stochastic optimization and Q-learning to dynamically change the thresholds. Protschky et al. use the Kalman filter and a generic statistical prediction model to perform traffic light control \cite{protschky2014adaptive}.

\section{System Model}
\begin{figure}[thpb]
	\centering
	\includegraphics[height=2.0in,width=3.5316in,angle=0]{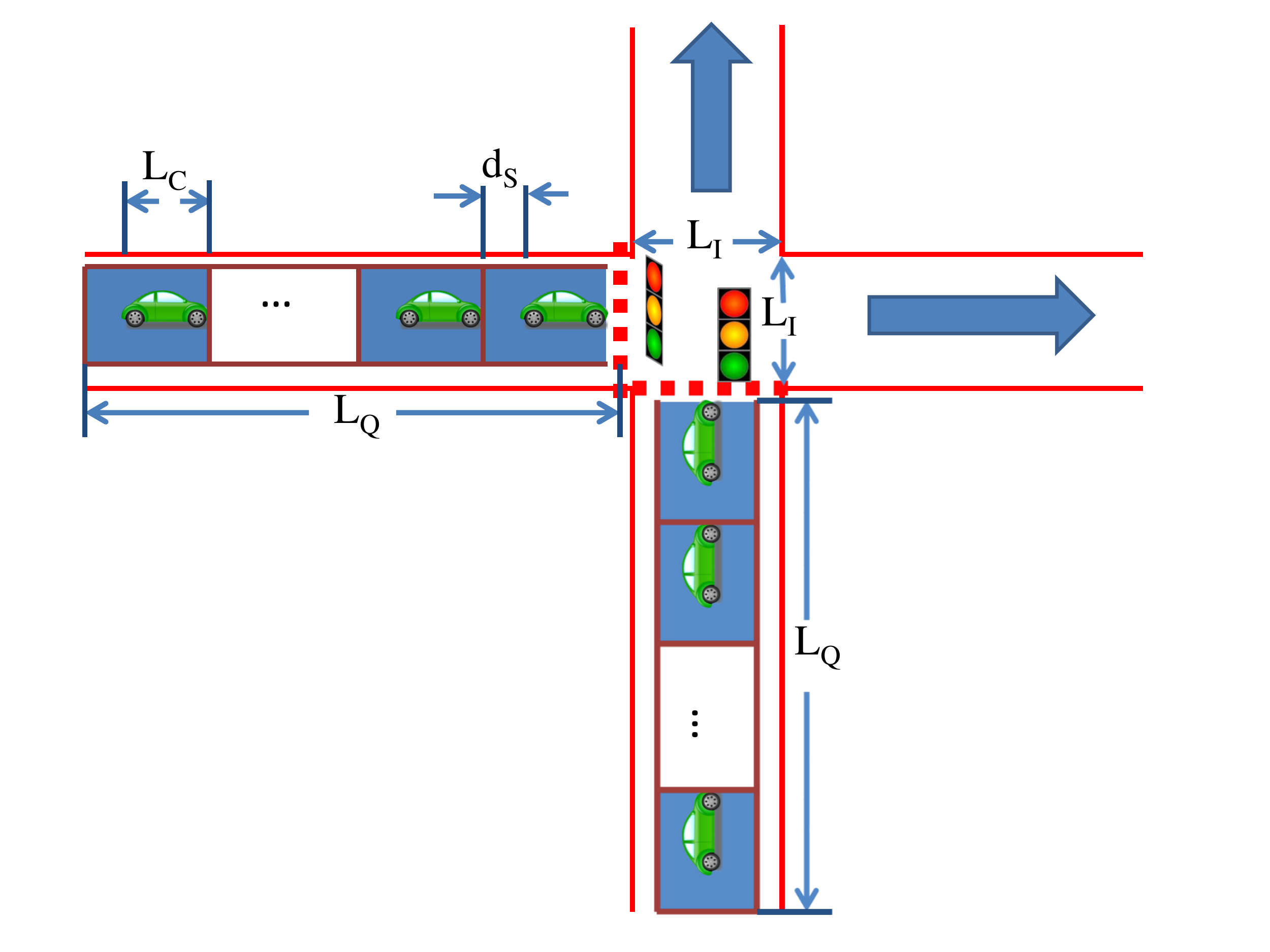}
	\caption{System Model Illustration}
	\label{system_model}
\end{figure}

We consider an isolated intersection of two one-way streets in this paper. Nonetheless, the results can be easily extended to more complex intersections. Fig. \ref{system_model} shows an illustration of the system model. In particular, we denote:

$L_{C}$: the length of the car. We assume that all the cars have equal length. 

$L_{I}$: the length of the intersection. We assume that the intersection has equal length in both directions. 

$L_{Q}$: the length of the vehicle queue. Assuming that each car is capable of communicating with the traffic light controller via wireless technologies, this is essentially the maximum distance of reliable wireless transmission.

$V_{max}$: the speed limit of the roads near the intersection. It is the maximum speed each car can have.

$v(t)$: velocity of a vehicle as a function of time.

$d_{S}$: the safe distance between two adjacent cars when they are both static. It is obvious that the safe distance depends on the velocities of the cars. Here, we are only concerned with the safe distance when both cars are static.

$V_{S}$: the safe velocity. When two adjacent cars are both initially static, $d_{S}$ meters away, and the first one starts to accelerate, it is the velocity of the first car at which the second one starts to accelerate. Here, we assume that the two cars are able to learn each other's information, including velocity, via wireless communications. Note that $V_{S} \le V_{max}$. We also would like to emphasize that this safe velocity can be translated into another safe distance by doing an integral, as shown later in the paper.

$N$: the maximum number of cars in each queue. For notation easy, we assume that $N= \frac{L_{Q}}{L_{C}+d_{S}}$ is an integer. This can be satisfied by simply choosing $L_{q}$ to be the integer multiple of $L_C+d_S$. 

$T_Y$: the duration of yellow light. We assume that $T_Y$ is greater than one second.

$D_{i}^j$: the amount of time the $i$-th car spends in the $j$-th queue. 

After introducing the system model of the intersection, we now discuss the car model. Since we are interested in providing QoS provisioning to the cars using the intersection, we need a simple model which can estimate the time it takes for the cars to travel through the intersection. We adopt a model used in \cite{xu2014communication}. Specifically, we assume that each car has the same mass $m$ and identical driving force $F$. The road friction $F_{1}(t)$ and wind dragging force $F_{2}(t)$ can be calculated as:
\begin{equation*}
F_{1}(t) =c_{1}v(t) \text{ and } F_{2}(t) = c_{2}v^{2}(t),
\end{equation*}
respectively, where $c_{1}$ and $c_{2}$ are constant coefficients. 
Applying Newton's Law, we obtain
\begin{equation}
m\dot v(t) + c_{1}v(t) + c_{2}v^{2}(t) = F \label{Newton}
\end{equation}
Because the vehicles are subject to $V_{max}$ near an intersection and are typically operated at low speed, we omit the wind's dragging force, which is only large when the velocity is high. (\ref{Newton}) can then be reduced to:
\begin{equation*}
m\dot v(t) + c_{1}v(t) = F \label{Newton_simplified}
\end{equation*}
Taking Laplace transform on both sides and assuming zero initial condition, we get the transfer function, which is first-order:
\begin{equation*}
G(s)=\frac{V(s)}{F(s)}= \frac{1}{ms+c_{1}}
\end{equation*}
The velocity $V(s)$ is:
\begin{equation*}
V(s)=\frac{F}{m}\frac{1}{s(s+\frac{c_{1}}{m})}
\end{equation*}
Let $K=F/c_{1}$ and $a=c_{1}/m$, the inverse Laplace Transform yields:
\begin{equation}
\label{v(t)}
v(t)=K(1-e^{-at}) 
\end{equation}%

\section {Main Results}
Our long term goal is to design an adaptive traffic light controller which is capable of providing QoS provisioning. In this paper, we first answer the following question: once a car enters a queue, what is the best and worst-case time that it will spend at the intersection? We use $D_{min}$ and $D_{max}$ to denote these two time, respectively. Only after they are determined, we can choose the right vehicle deadline $D$. 
$D_{min}$ corresponds to the best case scenario where a car enters a queue and travels through the intersection at velocity $V_{max}$. Therefore,
\begin{equation*}
D_{min}=\frac{L_{Q}+L_{I}}{V_{max}}.
\end{equation*}
Note that $D_{max}$ depends on the traffic light control algorithm. We emphasize that we are looking for the maximum delay introduced by the best scheduling algorithm. i.e., 
\begin{equation*}
D_{max}=\underset{\text{All traffic controllers}}{\min}\text{ } \underset{\text{All traffic conditions}}{\max}\text{  }{\max}(D_{i}^j).
\end{equation*}
Note that ${\max}(D_{i}^j)$ is the worst-case delay among all vehicles, and it depends on both the traffic condition and the traffic light controller. Obviously, it is maximized when the the traffic is extremely heavy in both directions so that even if the green light is constant on in one direction, the intersection will never idle.
To obtain $D_{max}$, we first establish the following result.
\begin{lem}
\label{travel_time_lemma}Suppose that $N$ cars are sitting in a queue and waiting for the green light. Assume that $L_{I}$ is wide enough for the first car to accelerate to $V_{max}$ before leaving the intersection. If the green light is turned on and stays on, it takes $T_{i}$ seconds for the $i$-th car to leave the intersection, $i\in \{1,\dots,N\}$:
\begin{align*}
T_{i}&=T_{i,w}+T_{i,a}+T_{i,t}, \text{ where}\\
T_{i,w}&=\frac{i-1}{a}[lnK-ln(K-V_{S})] \text{ is the time for the $i$-th car to reach the safe velocity $V_{S}$;}\\
T_{i,a}&=\frac{1}{a}[lnK-ln(K-V_{max})] \text{ is the time for the $i$-th car to accelerate to $V_{max}$; and}\\
T_{i,t}&=\frac{(i-1)(L_C+d_S)}{V_{max}}+\frac{L_{I}-(KT_{1,a}-\frac{V_{max}}{a})}{V_{max}} \text{ is the time for it to travel through the rest}\\
&\text{ of the intersection at maximum speed.}
\end{align*}
\end{lem}
\textbf{Proof:} We use induction to prove it. \\
\textit{Step 1:} We first focus on $T_{1}$, the time it takes the first car to depart. Because $T_{1,w}=0$, $T_{1}=T_{1,a}+T_{1,t}$. $T_{1,a}$ can be calculated using (\ref{v(t)}):
\begin{equation}
\label{accl_time}T_{1,a}=\frac{1}{a}[lnK-ln(K-V_{max})]
\end{equation}
The distance the first car travels during the transient period is:
\begin{equation}
\label{travel_distance}\int_{0}^{t}v(\tau)d\tau=Kt-\frac{K}{a}(1-e^{-at})
\end{equation}
Using (\ref{accl_time}) and (\ref{travel_distance}), the distance the first car travels from $0$ to $T_{1,a}$ is:
\begin{equation}
\int_{0}^{T_{1,a}}v(\tau)d\tau=KT_{1,a}-\frac{V_{max}}{a}
\end{equation}
Then, we calculate $T_{1,t}$:
\begin{equation}
\label{travel_time}T_{1,t}=\frac{L_{I}-(KT_{1,a}-\frac{V_{max}}{a})}{V_{max}}
\end{equation}

\textit{Step 2:} Assuming that $T_{i-1}$, $i\in\{1,\dots,N\}$, is known, we now figure out $T_{i}$. The difference between $T_{i}$ and $T_{i-1}$ has two parts. The first part is the time the $i$-th car has to wait for the $(i-1)$-th car's velocity to reach $V_{S}$. Let $\Delta T_w=T_{i,w}-T_{i-1,w}$, and using (\ref{v(t)}), we get
\begin{equation}
\label{delta_tw}\Delta T_w=\frac{1}{a}[lnK-ln(K-V_{S})]
\end{equation}
The second part is the time it takes to travel through the extra distance $L_C+d_S$ at velocity $V_{max}$. Let us use $\Delta T_t$ to denote this difference, and we have
\begin{equation}
\label{delta_tt}\Delta T_t=\frac{L_C+d_S}{V_{max}}
\end{equation}
Combining all the results above, we get:
\begin{align*}
T_{i,w}&=(i-1)\Delta T_{w}=\frac{i-1}{a}[lnK-ln(K-V_{S})]\\
T_{i,a}&=T_{1,a}=\frac{1}{a}[lnK-ln(K-V_{max})]\\
T_{i,t}&=(i-1)\Delta T_{t} +T_{1,t}\\
&=\frac{(i-1)(L_C+d_S)}{V_{max}}+\frac{L_{I}-(KT_{1,a}-\frac{V_{max}}{a})}{V_{max}} \text{ }\blacksquare
\end{align*}

We now use the results established in Lemma \ref{travel_time_lemma} to find $D_{max}$. It is obvious that $D_{max}$ is achieved when the traffic is heavy in both directions. Next, we will not give a precise definition of ``heavy"; instead, we give one scenario that will be guaranteed to yield $D_{max}$. Let us assume that the maximum service rate the intersection provides to the cars is $\mu  _{max}$; this is obtained when cars in one queue are traveling through the intersection at velocity $V_{max}$ with the right safe distance between adjacent ones. Then, the traffic condition can be considered as heavy or congested when the vehicle arrival rate to each queue is at least $\mu _{max}$ at all times. This traffic condition will lead to the maximum delay $D_{max}$.

\begin{lem}
\label{N_cars}
In heavily congested conditions, the optimal traffic control policy that yields $D_{max}$ is to let all $N$ cars in each queue pass the intersection during a single green light period.
\end{lem}
\textbf{Proof:} Because the traffic in both queues are heavy, there are $N$ static cars sitting in each queue when the traffic light is turned green for that queue.	Suppose that $M \in Z^{+}$ is the number of green light periods that are used to service $i$ cars, $i \in \{1,\dots,N\}$, i.e., there exists $n_{j} \in Z^+$ and $j \in \{1,\dots,M\}$ s.t. $\sum_{j=1}^{M} n_{j}=i$. We need to show that 
$T_{i} \le \sum_{j=1}^{M}T_{n_{j}}$. Using the results in Lemma \ref{travel_time_lemma}, we get:
\begin{align*}
T_{i}&=(i-1)\Delta T_{w}+T_{1,a}+(i-1)\Delta T_{t} +T_{1,t}\\
\sum_{j=1}^{M}T_{n_{j}}&=(i-M)\Delta T_{w}+MT_{1,a}+(i-M)\Delta T_{t}
+MT_{1,t}+(M-1)T_Y
\end{align*}
Subtracting $T_{i}$ from $\sum_{j=1}^{M}T_{n_{j}}$ yields:
\begin{equation}
\label{difference}
\sum_{j=1}^{M}T_{n_{j}}-T_{i}=-(M-1)\Delta T_{w}+(M-1)T_{1,a}\\
-(M-1)\Delta T_{t}+(M-1)T_{1,t}+(M-1)T_Y
\end{equation}
From (\ref{accl_time}), (\ref{delta_tw}), and $V_{max} \ge V_S$, we get
\begin{equation}
\label{T1a_greater}
T_{1,a} \ge \Delta T_w
\end{equation}
Because $\Delta T_t$ typically is less than $1$s and by assumption $T_Y$ is greater than $1$s, we have:
\begin{equation}
\label{delta_tt_smaller}
\Delta T_t \le T_Y
\end{equation}
Combining (\ref{difference}), (\ref{T1a_greater}), and (\ref{delta_tt_smaller}), we obtain: 
$T_{i} \le \sum_{j=1}^{M}T_{n_{j}} \text{ }\blacksquare$

\begin{lem}
\label{bad_luck}
Suppose that in heavily congested conditions, the optimal traffic control policy is used, i.e., $N$ cars in each queue are allowed to pass the intersection during a single green light period. In the $N$ cars sitting in the queue and waiting for the green light to be turned on,  the cars closer to the intersection spend not less time than those further away from the intersection.
\end{lem}
\textbf{Proof:} To prove this, we need to investigate when a car moves into a queue and when it leaves an intersection. In particular, we examine the behavior of a car outside of the queue when the green light is turned on for this queue. At that moment, there are $N$ cars sitting in the queue and $N$ cars sitting out of the queue. Invoking Lemma \ref{N_cars}, the $N$ cars in the queue will be serviced and then after the yellow light interval, and $N$ cars in the other queue will be serviced. Therefore, the $N$ cars outside of the queue will not be serviced until $2T_N+T_Y$ seconds later. 
Now, consider the $i$-th car outside of the queue. It takes $(N+i-1)\Delta T_w$ seconds for all the cars in front of it to accelerate to $V_S$. Let $T_{i,q}$ be the time it takes for the $i$-th car outside of the queue to enter the queue after it starts to accelerate. The total time for the $i$-th car to enter the queue after the green light is turned on is:
\begin{equation}
(N+i-1)\Delta T_w+T_{i,q}
\end{equation}
The amount of time the car spends in the queue before the next green cycle arrives is:
\begin{equation}
2T_N+2T_Y-(N+i-1)\Delta T_w-T_{i,q}
\end{equation}
The total time this car spends in the queue is:
\begin{equation}
2T_N+2T_Y-(N+i-1)\Delta T_w-T_{i,q} + T_i \label{delay_car_i}
\end{equation}
The difference of the time the ($i+1$)-th and the $i$-th car spend in the queue is:
\begin{align*}
\Delta_{i+1,i}&=-\Delta T_w + T_{i,q} -T_{i+1,q} + T_{i+1} - T_i \\
&=-\Delta T_w + T_{i,q} -T_{i+1,q} + \Delta T_w + \Delta T_t \\
&=-(T_{i+1,q}-T_{i,q}) + \Delta T_t
\end{align*}
Because $T_{i+1,q}-T_{i,q}$ is the time it takes to travel distance $L_C+d_S$ at average speed less or equal to $V_{max}$ and $\Delta T_t$ is the time it takes to travel the same distance at velocity $V_{max}$, we get:
$\Delta_{i+1,i} \le 0 \text{ } \blacksquare $

Lemma \ref{bad_luck} actually shows something very intuitive: in a queue packed with cars waiting for the traffic light to turn green, the cars closer to the intersection tend to be the unlucky ones that spend longer time in the queue. They would be luckier if they passed the intersection in the previous green light interval.

\begin{cor}
$D_{max}=2T_N+2T_Y-N\Delta T_w+ T_1$.
\end{cor}
\textbf{Proof}: $D_{max}$ can be obtained using Lemma \ref{bad_luck}, i.e., the maximum delay occurs to the first car in the queue waiting for the traffic light to turn green. Specifically, we let $i=1$ in (\ref{delay_car_i}) and get the closed form expression of $D_{max}$ shown above. $\blacksquare$

Since we now have $D_{min}$ and $D_{max}$, adaptive traffic light controllers can be built. Intuitively, the traffic controller should use the state of the system (number of existing cars in the queues), arrival rate, etc., to intelligently assign the deadline for each queue, based on which, the traffic lights are toggled between green and red. In this paper, we present a simple open-loop controller, which uses the theoretical results above to perform traffic light scheduling:
\begin{algorithm}
\caption{Adaptive open-loop traffic light controller}\label{euclid}
\begin{algorithmic}[1]
\State $Deadline \gets D_{max}$ 
\While{True}
\If {Intersection is not busy}
\State Turn on green light for the closest car
\Else
\If {The deadline of any car waiting for the green light is about to be missed}
\State Toggle the traffic lights
\EndIf
\EndIf
\EndWhile
\end{algorithmic}
\end{algorithm}
\section{Simulation Results}
We run simulation to verify the theoretical results and the performance of the above adaptive traffic controller. In our experiments, we let the numbers of vehicles in the two queues be $M_1$ and $M_2$, respectively. Initially, all vehicles are at least $L_Q$ meters away from the intersection, and the distance between two adjacent cars in queue $2$ is uniformly distributed between an interval $[d_1, d_2]$. The $M_1$ cars in queue 1 are uniformly distributed between the locations of the first and the last cars in queue 2. The initial velocities of all cars are uniformly distributed between $0$ and $V_{max}$, which is set to $13.3m/s$, i.e., $30$ miles per hour. Other parameters' values are: $m=1500kg$, $F=44444N$, $c_{1}=1000$, $L_C=4m$, $L_I=25m$, $N=20$, $d_S=1$, $L_Q=100m$, and $T_Y=3s$. Using these values, we calculate that $D_{min}=9.40s$ and $D_{max}=31.05s$.

Traffic congestion depends on the vehicle arrival rate or car density in \textit{each} queue: if one queue has zero or very few cars, then it is likely that there is no traffic jam at the intersection, provided that an adaptive traffic control mechanism is in place. For this reason, we define $R$, the ratio between the numbers of cars: $R=M_{1}/M_{2}$. In our simulation, $M_1\le M_2$ and $M_2$ is fixed at 200. Therefore, $R\le 1$. Note that traffic congestion depends on not only $R$, but also the values of $d_1$ and $d_2$.

We compare the performance of the adaptive controller specified in Algorithm 1 with two other controllers:\\
\textit{Fixed-cycle controller 1}: the green light and the red light cycles of both queues are $T_N$ seconds.\\
\textit{Fixed-cycle controller 2}: the green and the red light cycles of queue 2 are $T_N$ and $R \times T_N$, respectively.

\begin{figure}[thpb]
\centering
\includegraphics[height=2.5in,width=3.5in,angle=0]{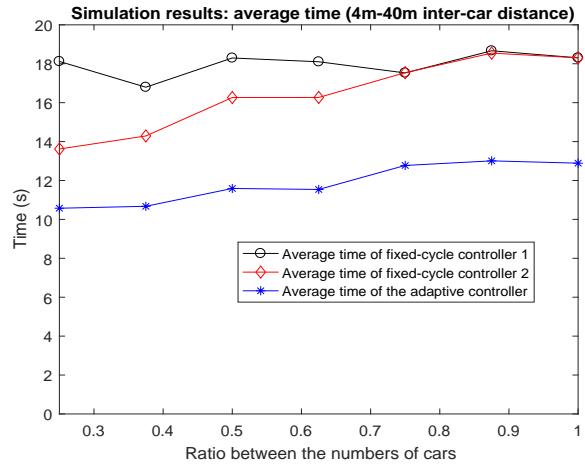}
\caption{Average time a car spends at the intersection (4m-40m inter-car distance)}
\label{avg_40m}
\end{figure}

\begin{figure}[thpb]
\centering
\includegraphics[height=2.5in,width=3.5in,angle=0]{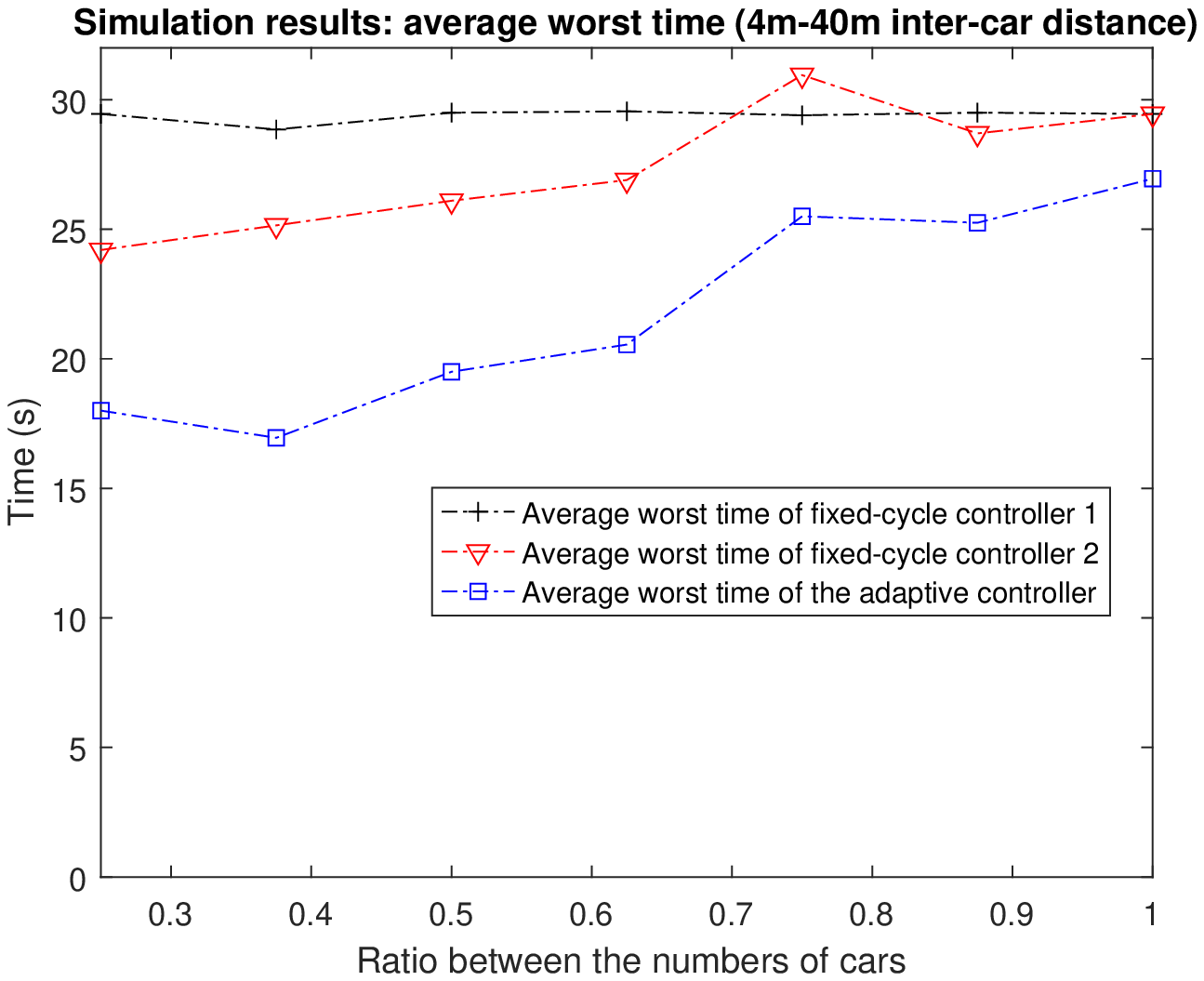}
\caption{Average of the worst time in each queue (4m-40m inter-car distance)}
\label{worst_40m}
\end{figure}

\begin{figure}[thpb]
\centering
\includegraphics[height=2.5in,width=3.5in,angle=0]{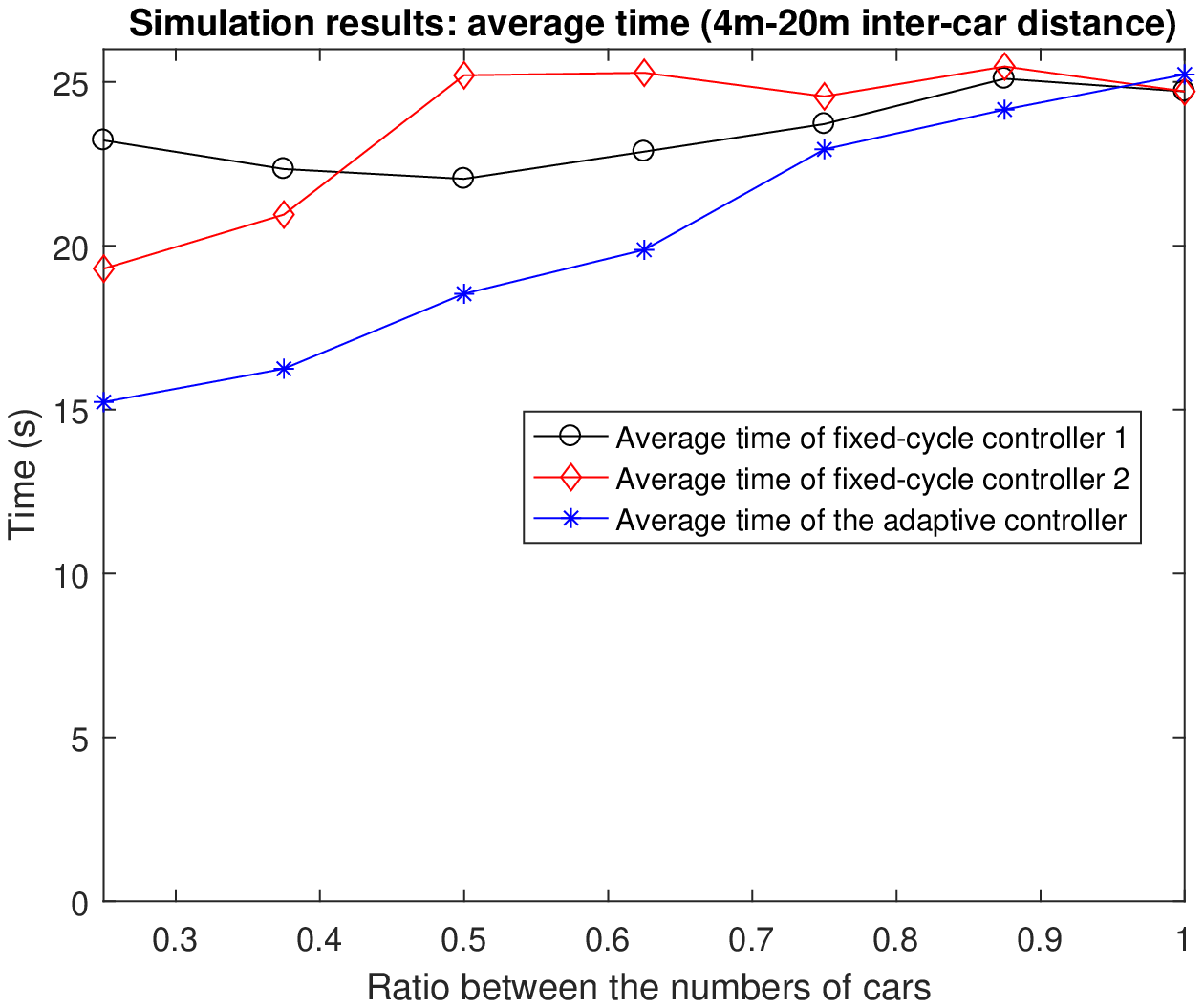}
\caption{Average time a car spends at the intersection (4m-20m inter-car distance)}
\label{avg_20m}
\end{figure}

\begin{figure}[thpb]
\centering
\includegraphics[height=2.5in,width=3.5in,angle=0]{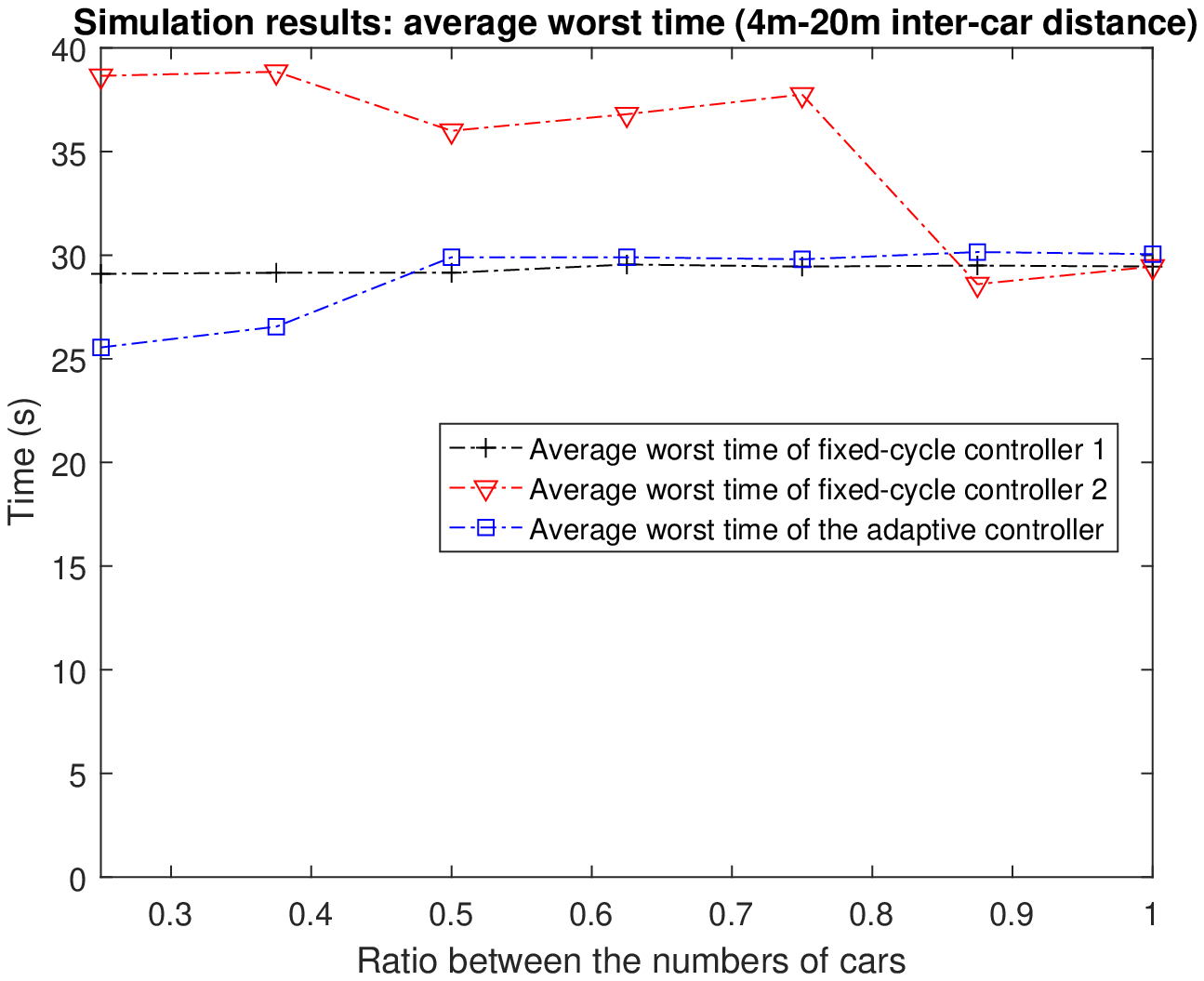}
\caption{Average of the worst time in each queue (4m-20m inter-car distance)}
\label{worst_20m}
\end{figure}

Fig. \ref{avg_40m} and Fig. \ref{worst_40m} show the results when $d_1=4m$ and $d_2=40m$. In this case, the traffic is light, and the adaptive controller we propose reduces the average time by a large percentage (around $40\%$ when $R=0.25$ in Fig. \ref{avg_40m}). The reason why the large deadline ($31.05s)$ does not affect things is that when the traffic is light, the intersection tends to become idle long before a waiting car's system time reaches this deadline. This will turn on the green light for the cars waiting in the queue. Another observation is that fixed-cycle controller $2$ is better than controller $1$. However, it is not as good as the adaptive controller.

Nonetheless, the large deadline does help in the heavy traffic scenario, as shown in Fig. \ref{avg_20m} and Fig. \ref{worst_20m}, where $d_1=4m$ and $d_2=20m$. In this case, it is less likly for the intersection to be idle, as $R$ gets closer to $1$. As a result, the large deadline used in the adaptive controller prevents the traffic light from toggling too often; this, in turn, helps with lowering the averge time. We also point out that when $R$ approaches $1$, there is not much optimization to be done since the optimal scheduling is simply divide the green light time equally between the two queues, as indicated in Lemma \ref{N_cars}. In addition, when the traffic is moderate ($R\le 0.75$), fixed-cycle controller $2$ is worse than controller 1, indicating that simply modifying the cycle length of fixed-cycle scheduling does not always help.

Overall, the adaptive controller significantly outperforms the other two fixed-cycle ones when the traffic is not very heavy and maintains roughly the same performance when the traffic is extremely heavy.

\section{Conclusions and Future Work}
In this paper, we model the process of cars passing through an intersection and come up with the shortest and the longest time they could possibly spend at the intersection. The results are significant because they help people better understand how traffic light controllers can be designed in order to provide quality-of-service (QoS) to travellers in intelligent transportation systems. We also propose a very simple adaptive controller, which is shown to be much better than two other fixed-cycle ones in most scenarios.

Our future work involves improving the adaptive controller and providing QoS guarantee (e.g., the worst-case waiting time at an intersection) to each individual car. It has been shown in our simulation that when the traffic is light and the controller is adaptive, it is possible to improve the the average time and worst time of both queues. However, we have also seen that when the traffic is no longer light, the improvement of wait time over one queue actually hurts the wait time of the other. We are interested in studying how the social aspects affect the decision making in adaptive traffic light control.

\end{document}